\documentclass[journal=jacsat,manuscript=article]{achemso}
\usepackage[version=3]{mhchem} 
\usepackage[caption=false]{subfig}
\usepackage{color}
\usepackage{soul}
\usepackage{textcomp}

\author{Vivekanand Shukla}
\email{Vivekanand.Shukla@physics.uu.se}
\author{Anton Grigoriev}
\email{Anton.Grogoriev@physics.uu.se}
\author{Rajeev Ahuja}
\affiliation[Uppsala University]
{Condensed Matter Theory Group, Materials Theory Division, Department of Physics and Astronomy, Uppsala University, Box 516, SE-75120, Uppsala Sweden}
\alsoaffiliation[Second University]
{Applied Materials Physics, Department of Materials and Engineering, Royal Institute of Technology (KTH), SE-10044, Stockholm Sweden}

\title[An \textsf{achemso} demo]
{Electronic transport properties in 90\textdegree rotated bilayer black phosphorus nanojunction: A first principle investigation }

\clearpage

\title[An \textsf{achemso} demo]
{Rectifying properties in 90\textdegree rotated bilayer black phosphorus nanojunction: A first principle study }

\abbreviations{IR,NMR,UV}


\begin{document}

\begin{tocentry}
\includegraphics[width = 4.5cm,height=3.6cm]{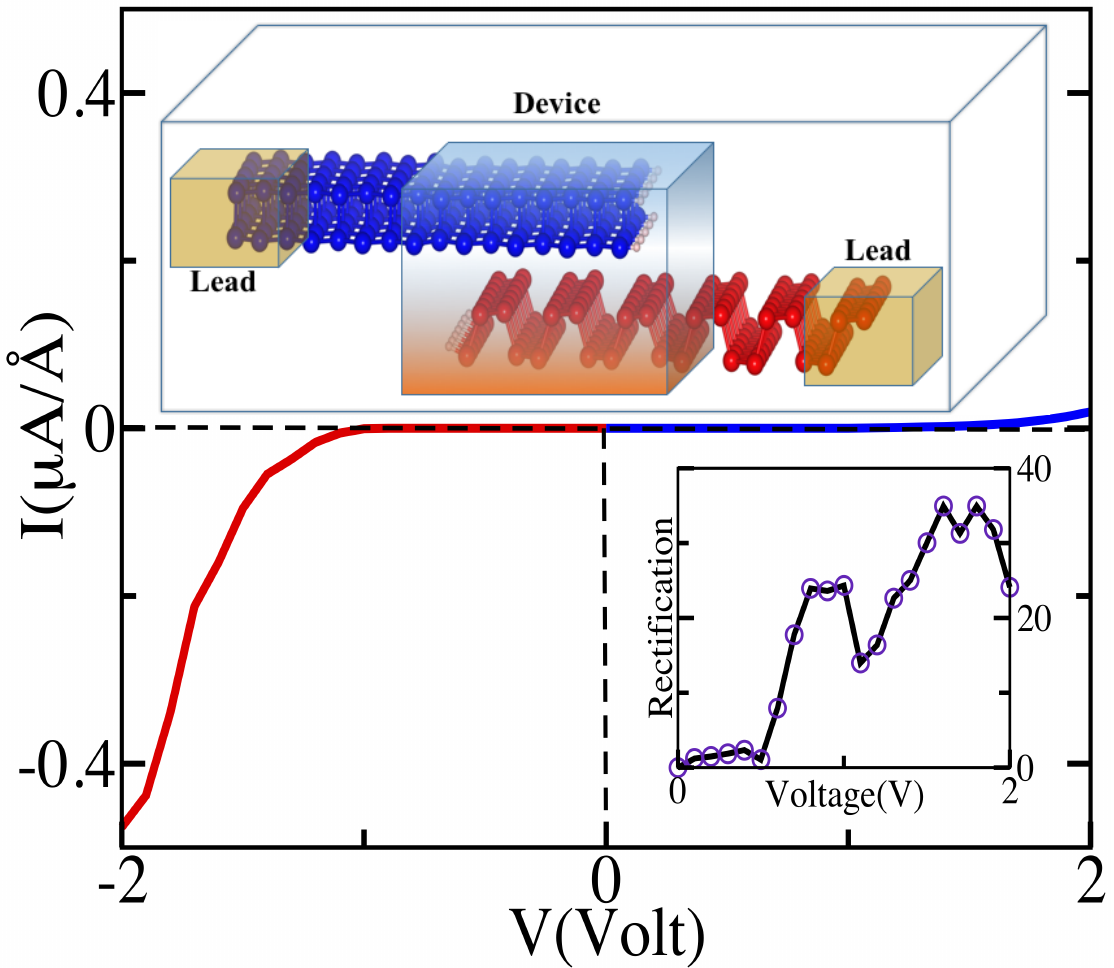}
\end{tocentry}

\begin{abstract}
We explore the possibility of using van dar Waals bonded heterostructures of stacked together 2D bilayer black phosphorus (BP) for nanoscale device applications. The electronic property of BP in AA stacking and 90\textdegree twisted is studied with density functional theory. Further, we study the homogeneous nanojunction architecture of BP to use its anisotropic properties. Using the first principle simulations along with NEGF approach, we calculate quantum transport properties of the nanojunction setup. The interlayer directionally dependent current characteristics are explained in different setups. Our result revealed that 90\textdegree twisted nanojucntion device would be a potential rectifier despite having no p-n junction characteristic only due to the intrinsic anisotropy of the material, making tunneling between armchair- and zigzag-directional BP sheets asymmetric. 
 
\end{abstract}

\section{Introduction}
Two dimensional (2D) materials unveil wide range of new physical phenomena, which would be elusive in their bulk counterparts, mass-less Dirac fermions in graphene are a good example\cite{geim2007rise,zhang2005experimental,xu2014spin}. Apart from the fundamental interest in 2D materials, intensive research is conducted towards their real world applications, like electrochemical use, energy storage and so on\cite{gholamvand2016electrochemical,shukla2017curious,jena2017borophane,shukla2018borophene}. One of the important areas is in electronic devices, where significant exertion is being devoted to integrate 2D materials into nano and micro electronic devices in the forms other than monolayer\cite{frisenda2018atomically,novoselov20162d}. Synthesis of new materials by stacking monolayers of same or two different materials, t.ex. the epitaxial growth of superlattices, has been a subject of fervent research for decades.\cite{esaki1970superlattice} The concept of van der Waals (vdW) heterostructures was conceptualized by Geim and Grigorieva a few years back.\cite{geim_van_2013} These vdW heterostructures are materialized by stacking of two 2D crystalline atomic layers, with no chemical bond between them\cite{pontes2018layer,padilha2015van}. The stacking patterns unfold the world of possibility of any desired heterostructure with specified chemical/physical and electronic property. These hetersotructures have been studied widely both theoretically and experimentally like graphene/hexagonal boron nitride (G/h-BN), MoS$_2$/G, MoS$_2$/WSe$_2$ for band alignment and charge transfer which makes them suitable for optoelectronic applications\citep{iqbal2018gate,pierucci2016band,komsa2013electronic}. WSe$_2$/h-BN/G heterostructure\citep{Li2017TwodimensionalNP} was reported for programmable p-n junction diodes and the same property has also been recited for TMD/G, h-BN/TMD, G/BN and TMD/TMD and several others\citep{khan2018high,choi2014lateral,chen2018gate,jin2017interlayer,lee2014atomically}.

	Recent successful exfoliation of a new class of two-dimensional materials, layered black phosphorus (BP) or phosphorene, has drawn both theoretical and experimental attention owing to its conspicuous electronic properties and high charge carrier mobilities of the resulting layered material.\cite{li2014black,churchill2014two,liu2014phosphorene} Monolayer BP have extremely high hole mobility (1000 cm$^2$V$^{-1}$S$^{-1}$) and on/off ratio of $10^4$,\cite{liu2014phosphorene,li2014black} which compares better to MoS$_2$\citep{lin2012mobility}. Application of strain and electric field are helpful in tuning electronic properties of BP\cite{rodin2014strain,fei2014strain,ccakir2014tuning,koenig2014electric}. BP has also been investigated in heterostructure form, Graphene-BP (G-BP) heterstructure has been reported for Schottky barrier tuning with the application of electrostatic gating.\cite{padilha2015van} BP-TMCs\cite{deng2014black,huang2015electric} and Blue phosphorus/Black phosphorus heterostructres\cite{huang2016tunable} are also reported for the tunable electronic properties. For the homogeneous device, few layers of BP are predicted to have good tuneability and richer electronic properties compared to a monolayer, such as layer dependent band gap and carrier mobility\cite{qiao2014high,ccakir2015significant,tran2014layer}. The predicted layer dependent bandgap of few layer BP has been confirmed experimentally by Das et al., evidenced by layer dependent transport\cite{das2014tunable}. Theoretically the bilayer BP has been reported in different stacking styles like AA, AB, AC and A-delta\cite{ccakir2015significant, lei2016stacking}. Its bandgap highly depends upon these stacking. However, there is also possibilities of stacking two layer of BP at 90\textdegree which gives rise the isotropic properties in terms of carrier transport and electronic structure. This twisted bilayer has already been reported for tunable anisotropy with electrostatic field and strain effect\cite{sevik2017gate,cao2016gate,xie2018strain}. Experimentally, BP mono- or few-layer form with controlled vertical gate voltage was used in device configuration as reported by Ameen et. al.\cite{ameen_few-layer_2016}

	Geim and Grigorieva introduced the concept of vertical vs. lateral junctions. However, in practical applications, most of the 2D hetero and homo junctions are neither purely vertical nor completely lateral as is often assumed in theory\citep{zhang2016band,yu2016carrier}. Nanodevices take form of monolayer-bilayer-monolayer junction (ML-BL-ML) structure. Ponomarev \textit{ et. al. } has reported WSe$_2$/MoSe$_2$ and WSe$_2$/MoS$_2$ nanojunctions in the form of ML-BL-ML for diode applications in experiment\cite{ponomarev2018semiconducting}. Graphene and Carbon nanotubes (CNT) has also been reported in these ML-BL-ML form for negative differential resistance behavior\citep{liu2011negative,leech2018negative}.

\hspace{2cm} In this work, we performed the density functional theory (DFT) with non-equilibrium Green’s function (NEGF) calculations to model the bilayer structures and ML-BL-ML form of nanojunction setup of BP. We started with AA stacked BP bilayer and compared it with 90\textdegree twisted bilayer in 5x7 supercell. After having insight of structural and electronic properties of these bilayers, we focused our study in ML-BL-ML architecture of nanodevice. Using the NEGF along with the Landauer formula we calculated the transmission function and studied I-V characteristics of the devices. Interestingly we report the rectifying behavior in twisted bilayer devices. 

\section{Results and discussion}
As a first step to use BP bilayer in nano-jucntion devices architecture, we simulated AA stacked bilayer and studied electronic and structural properties. We simulated different stacking of the bilayer BP like AA, AB, and AC. Generally, the AB stacking is natural stacking with highest possibility for from but in our work, we consider AA stacking for further use to facilitate comparison with the twisted bilayer structure. The calculated lattice parameter for AA stacking is as 4.50\AA{} and 3.4\AA{} in the armchair and zigzag direction which clearly relates with previously reported results\citep{ccakir2015significant}. All these stacking patterns show the direct band gap properties. These stackings can also tune optical properties dramatically which has been reported by \citep{ccakir2015significant}. The AA-stacked bilayer structure is shown in Fig. 1(a) where, the interlayer distance is calculated as 3.3\AA{}. Fig. 1(b), depicts the band structure of AA-stacked bilayer BP, which demonstrates that the anisotropy exists in the armchair and zigzag directions and the calculated bandgap is 0.37 eV. This anisotropy and its effect on electron and hole mobility can be found elsewhere\citep{fei2014strain,qiao2014high}. In the band structure picture, we see no degeneracy in conduction band minimum (CBM) and valence band maximum (VBM). This owes to strong interlayer interaction of the two extra lone pair electrons in interlayer phosphorus atoms. Further, this strong layered coupling is also responsible for lowering the band gap in few-layer BP than the monolayer (reported in supporting information). In the bilayer form, BP has the direct band gap properties which makes bilayer BP attractive material for the electronic application.

	Along with the symmetric stacking scheme of bilayer BP, there is also the possibility to twist the bilayer by some rotational angles, which creates Morrie pattern\citep{kang2017moire}. Here we twisted the upper layer by 90\textdegree, which in some sense makes the system laterally symmetric in structure. Fig 1(c) manifests the twisted bilayer structure, where the zigzag direction in the upper layer is the same as the armchair direction in the lower layer. Supercell approach has been used by taking the lateral dimension as 5x7 times that of the unit cell to minimize the lattice mismatch less than 1\%. In this supercell, the two bilayers (one in the supercell and its replication) are at the distance of 30\AA{}, which was selected to exclude the interaction in the vertical direction from consideration. The calculated lattice parameter for this twisted bilayer unit cell is of 23.09\AA{} and 23.00\AA{} and the interlayer spacing in this fully relaxed 90\textdegree twisted bilayer cell is calculated as 3.4\AA{}, which is 0.1\AA{} higher than that of the AA-stacked bilayer. We used vDW D2 correction for the bilayer unit cell and supercell calculations\citep{harl2010assessing}. Full relaxation of the cell volume along with the atomic positions was conducted. P-P bond in the twisted bilayer shows only the subtle changes in comparison to naturally AA-stacked bilayer in previous sections. 

\begin{figure}[h]
 \centering
\subfloat[]{\includegraphics[width = 3.5cm]{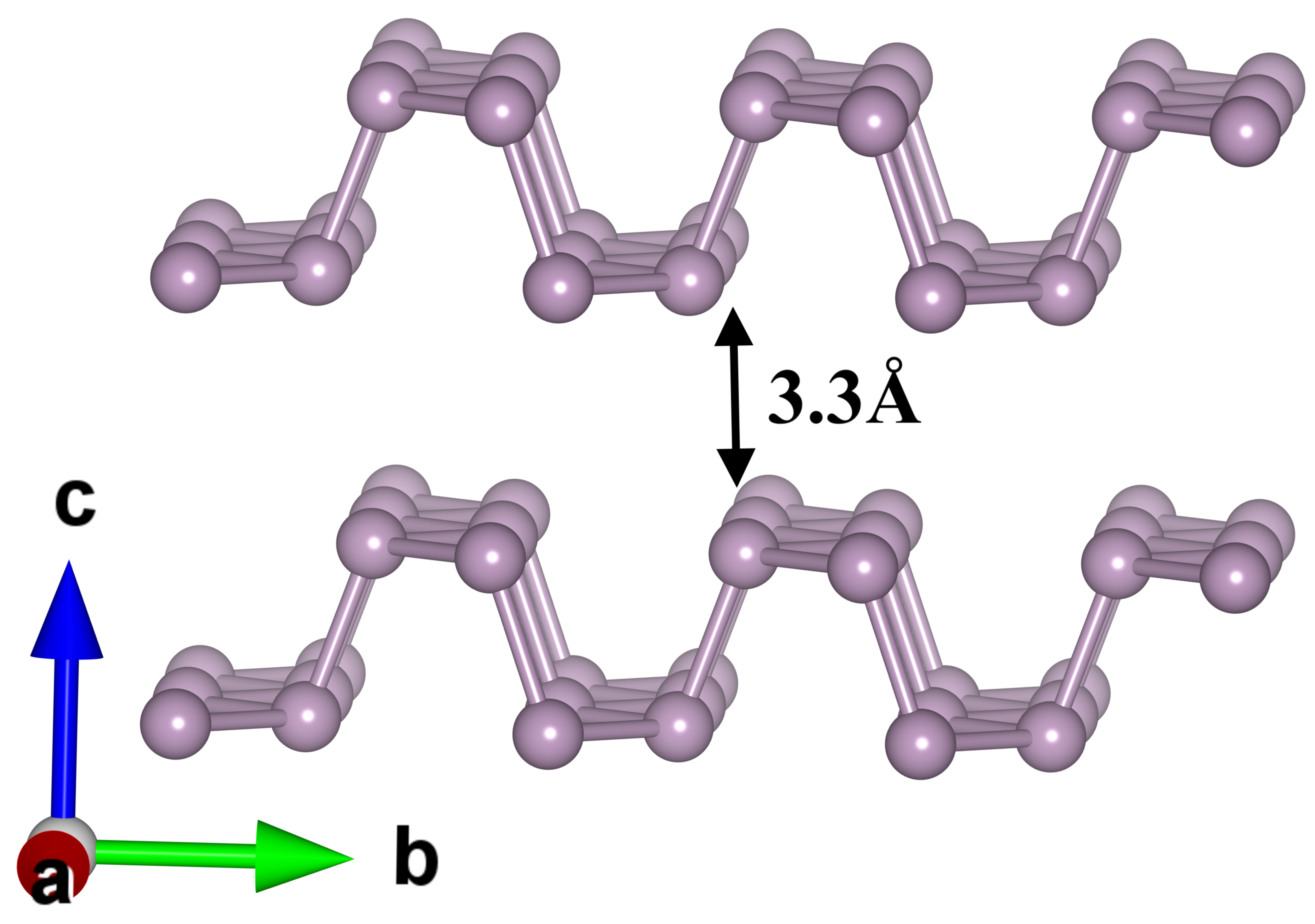}}
\subfloat[]{\includegraphics[height=4cm]{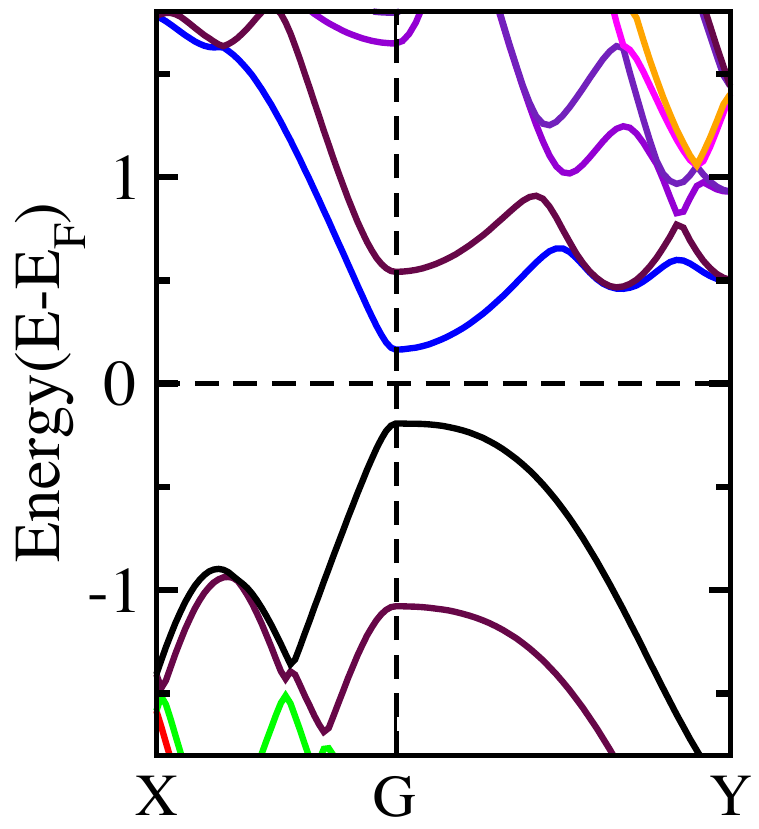}}
\subfloat[]{\includegraphics[width = 4cm]{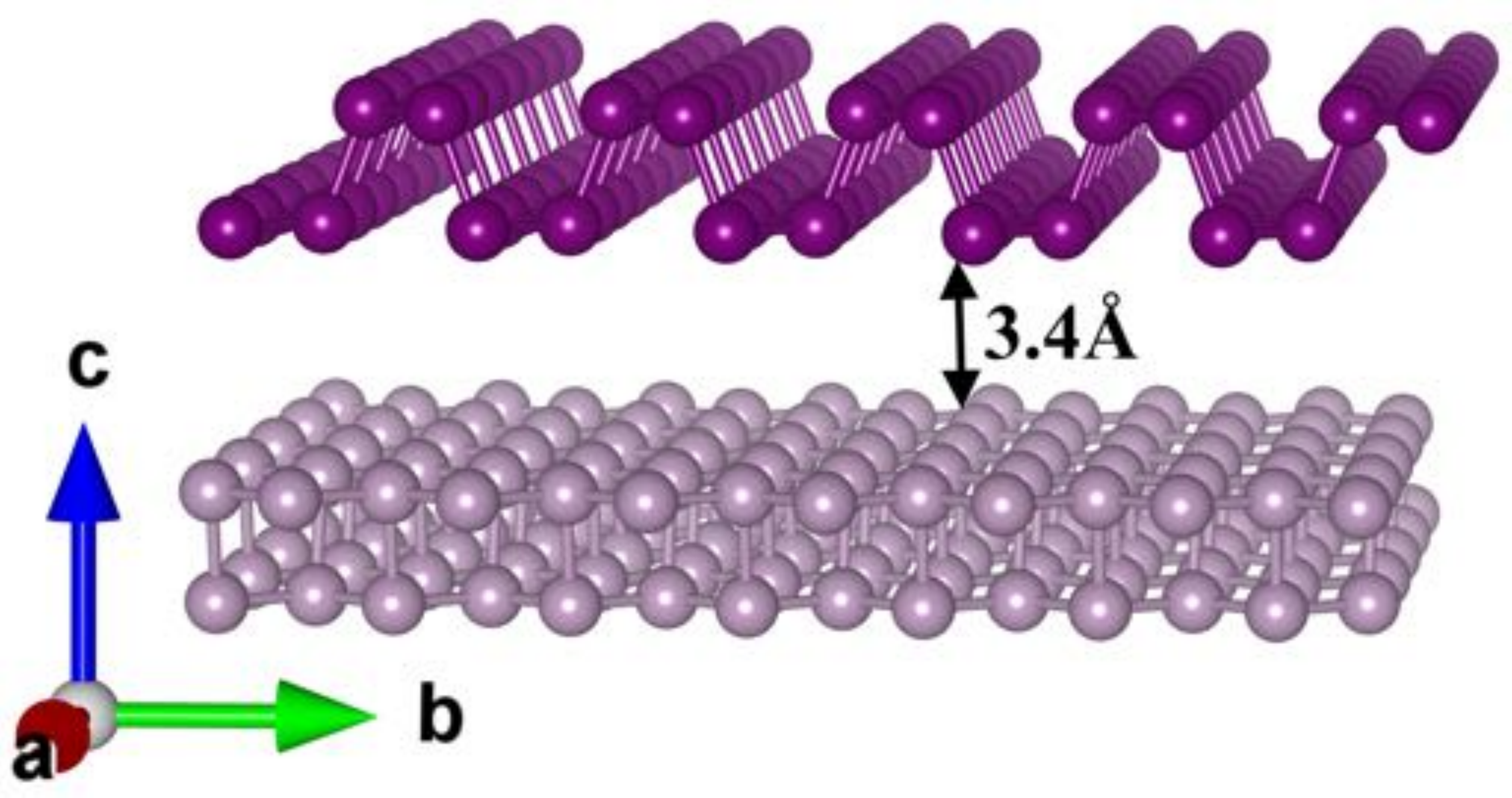}}
\subfloat[]{\includegraphics[height=4cm]{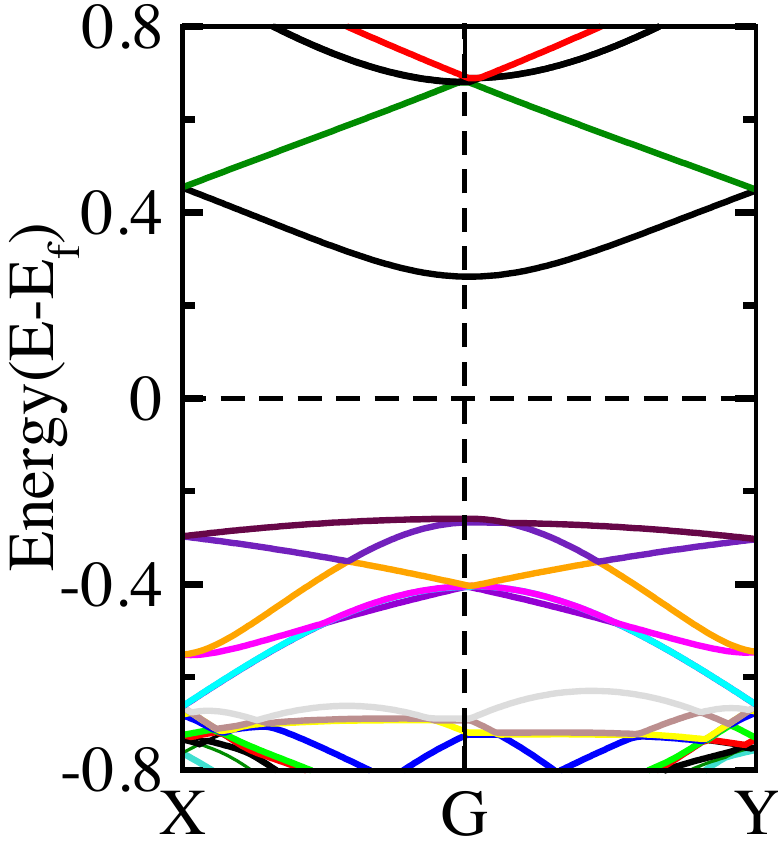}}
\captionsetup[figure]{font=small,skip=0pt}
\caption{(a) Schematic picture of AA stacked bilayer (b) Band structure for bilayer (c) 90\textdegree twisted stacked bilayer structure (Upper layer in twisted by 90\textdegree to the lower layer) (d) Band structure for twisted stacked bilayer (brillouin zone for the twisted bilayer is square)}
\label{fig1}
\end{figure}

	Reciprocal space geometry for the twisted bilayer would be square in the Brillouin zone. In contrast to the AA-stacked bilayer, 90\textdegree twisted bilayer has an x-y isotropic electronic structure, means the energy band around X-G and G-Y are symmetric in Fig. \ref{fig1}(d). Calculated band gap for twisted bilayer is of 0.48 eV, which is bigger than that of the AA-stacked bilayer. This clearly corresponds to the lower interaction in between the layers, which is also supported by increased interlayer spacing. CBM is single degenerate whereas the VBM is found to be doubly degenerate by the value of 0.01 eV. Degeneracy in VBM advocates the of lack of significant hybridization near the VBM whereas in CBM states are strongly coupled. These degenerate states are separated either localized in the top layer or localized in the bottom layer. More than that, this degeneracy paves the isotropy in the system where valence band in X-G is defined by the upper layer and G-Y is defined by bottom layer. This degeneracy and its dependence on the vertical electric field and external strain has been explained previous reports. The vertical electric field in this twisted bilayer can induce the anisotropy in the system that can control the transport properties of this bilayer in the device form. The vertical electric field tune the anisotropy in either direction of bands, if electric field applied from top then band G-X band will get tuned and when it applied from down then G-Y band can get tuned (Supporting information). External strain also leads to the effective anisotropy in this system reported by Xie \textit{et. al.} \cite{xie2018strain}. Further, Fig. S1 shows the projected density of states (pDOS) plot for this twisted bilayer. The projected density in the respective layer illustrates that they do not differ with each other, which clearly demonstrates that there is no charge transfer in between the layers unlike the heterostructure made from two different materials, for example, G/BP, G/MoS$_2$, G/h-BN, MoS$_2$-BP and even in Black Phosphorus-Blue Phosphorus. This is a homogeneous bilayer and only symmetric twisting does not give rise the charge transfer process.

\begin{figure}
\centering
\subfloat[]{\includegraphics[height=2.45cm,width = 8.5cm]{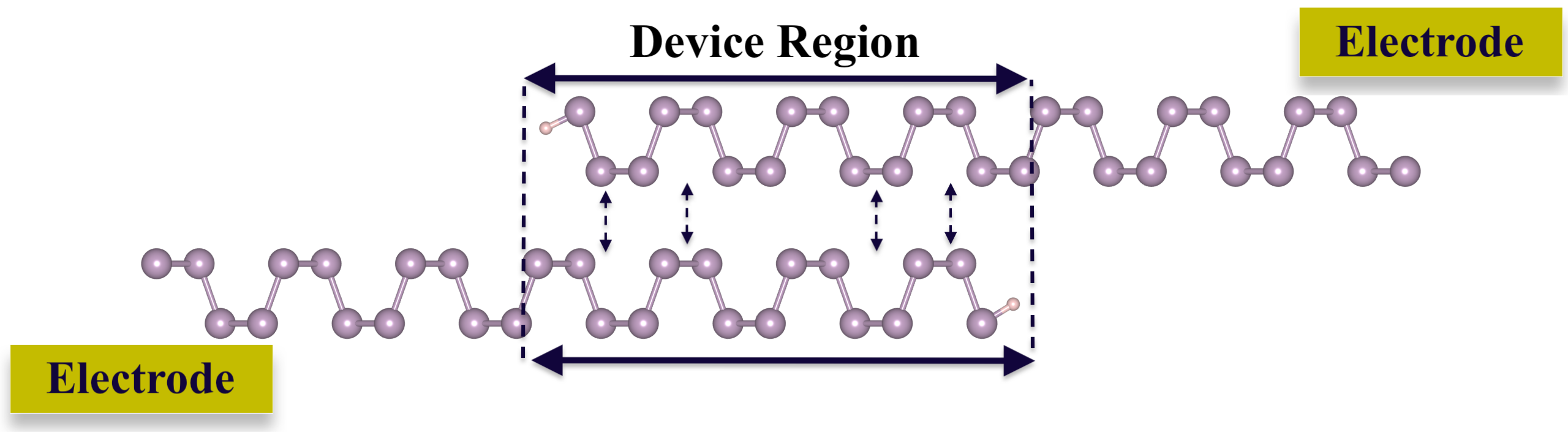}}
\subfloat[]{\includegraphics[height=2.45cm,width = 8.5cm]{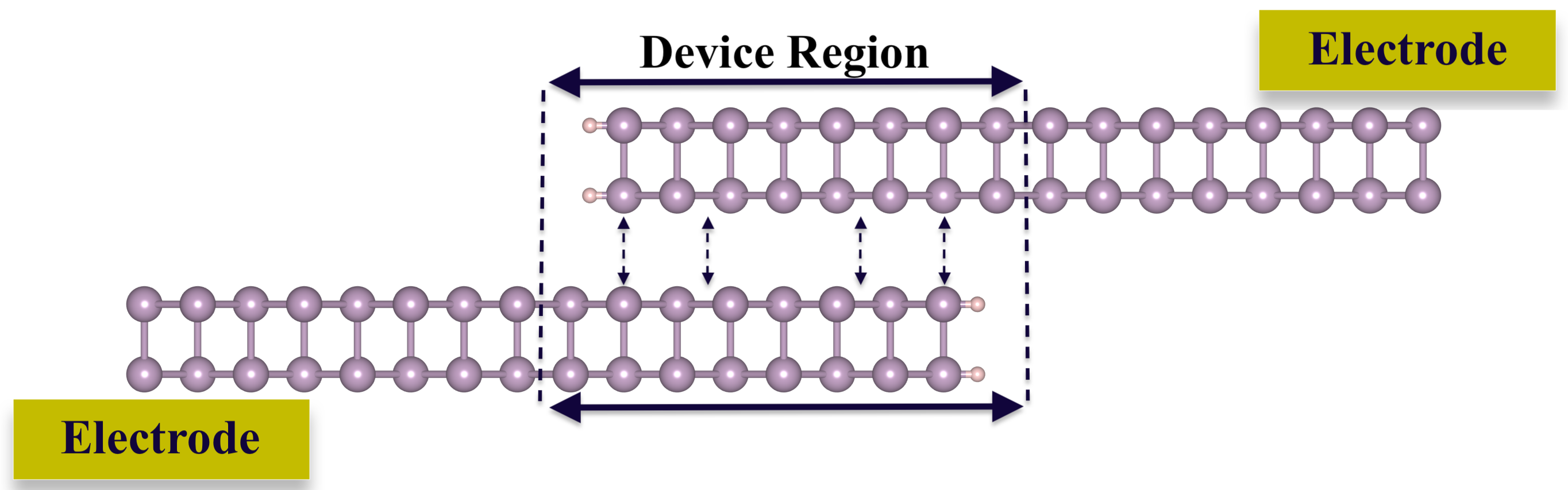}}\quad
\captionsetup[figure]{font=small,skip=0pt}
\caption{Schematic pictures for ML-BL-ML (Junction area has bilayer) nanojunction device set in (a) Armchair direction (b) Zigzag direction}
\end{figure}

	After having a look in the electronic and structural properties of the two different bilayers we introduce the homogeneous nanojunction. Here the two layers of BP in the form of AA stacking were put on each other and then in either direction, dangling bonds were passivated by hydrogen atoms. It gives the ML-BL-ML architecture, where the central part has a bilayer of the armchair in fig. 2(a) and zigzag in fig. 2(b). Monolayers in these ML-BL-ML junctions are armchair and zigzag form respectively. In our case, we deliberately created this junction to understand the difference of direction dependence in this kind of nanojunctions of BP. In these two cases, a potential well is formed in this scattering region (bilayer region). The similarity between these two nanojunctions is that the electron has to pass through the vertically stacked area through interlayer tunneling. Transport properties would be estimated from nearly the same length of the scattering region in these two devices. 

	We would first investigate the armchair direction and further will take the consideration of the zigzag one. Zero bias transmission in armchair nanojunction is shown in Fig. 4(b), where it shows the transmission gap of 0.9 eV. This gap resembles the monolayer electrodes as the electron are injected through the monolayer of BP and the bilayer (nanojunction) region behaves like the scattering owes to the broken periodicity. This scattering center is like a quantum well where electron can resonate and then pass through the tunneling mechanism. 
 
\begin{figure}
\subfloat[]{\includegraphics[width = 5cm]{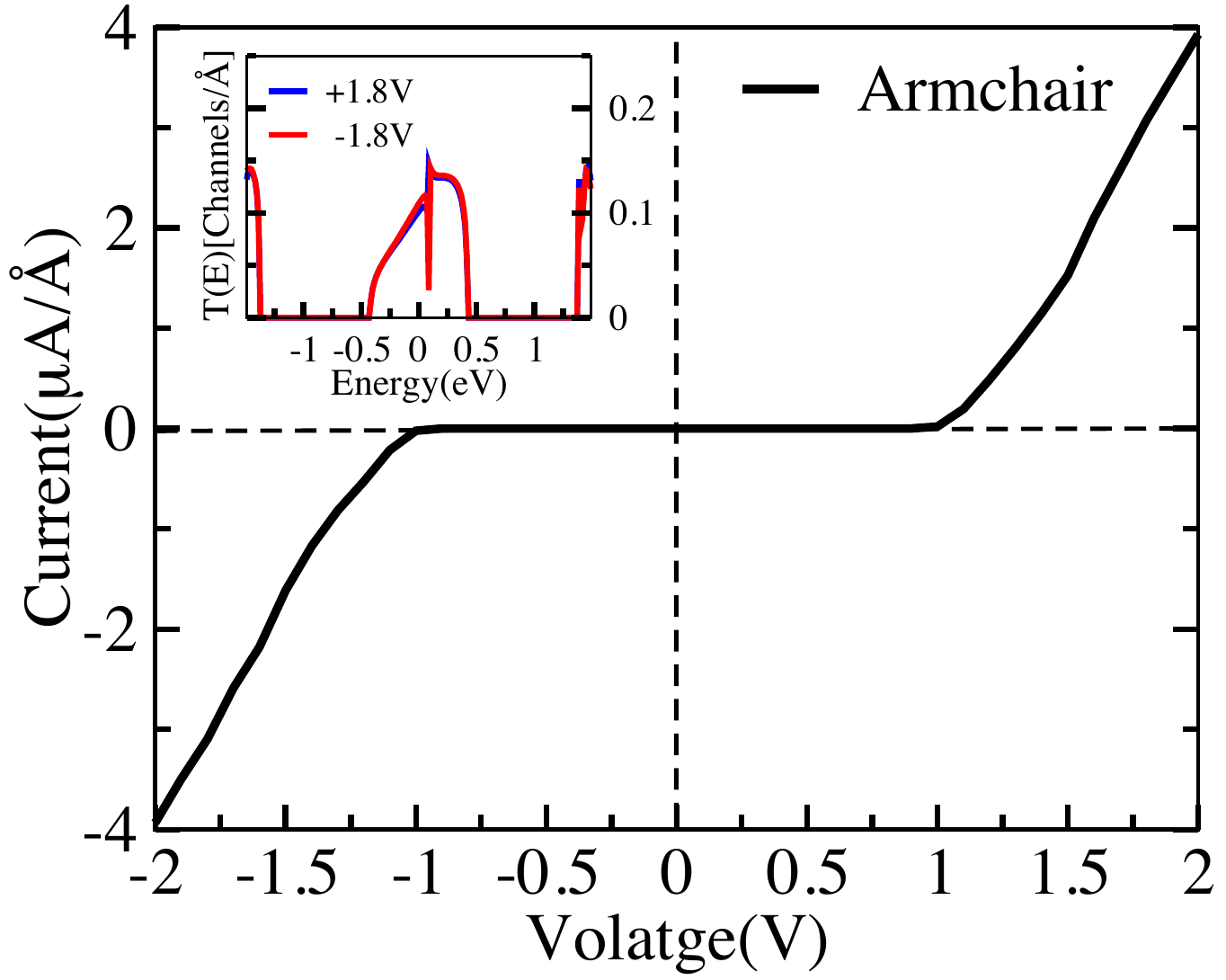}}
\subfloat[]{\includegraphics[width = 5cm]{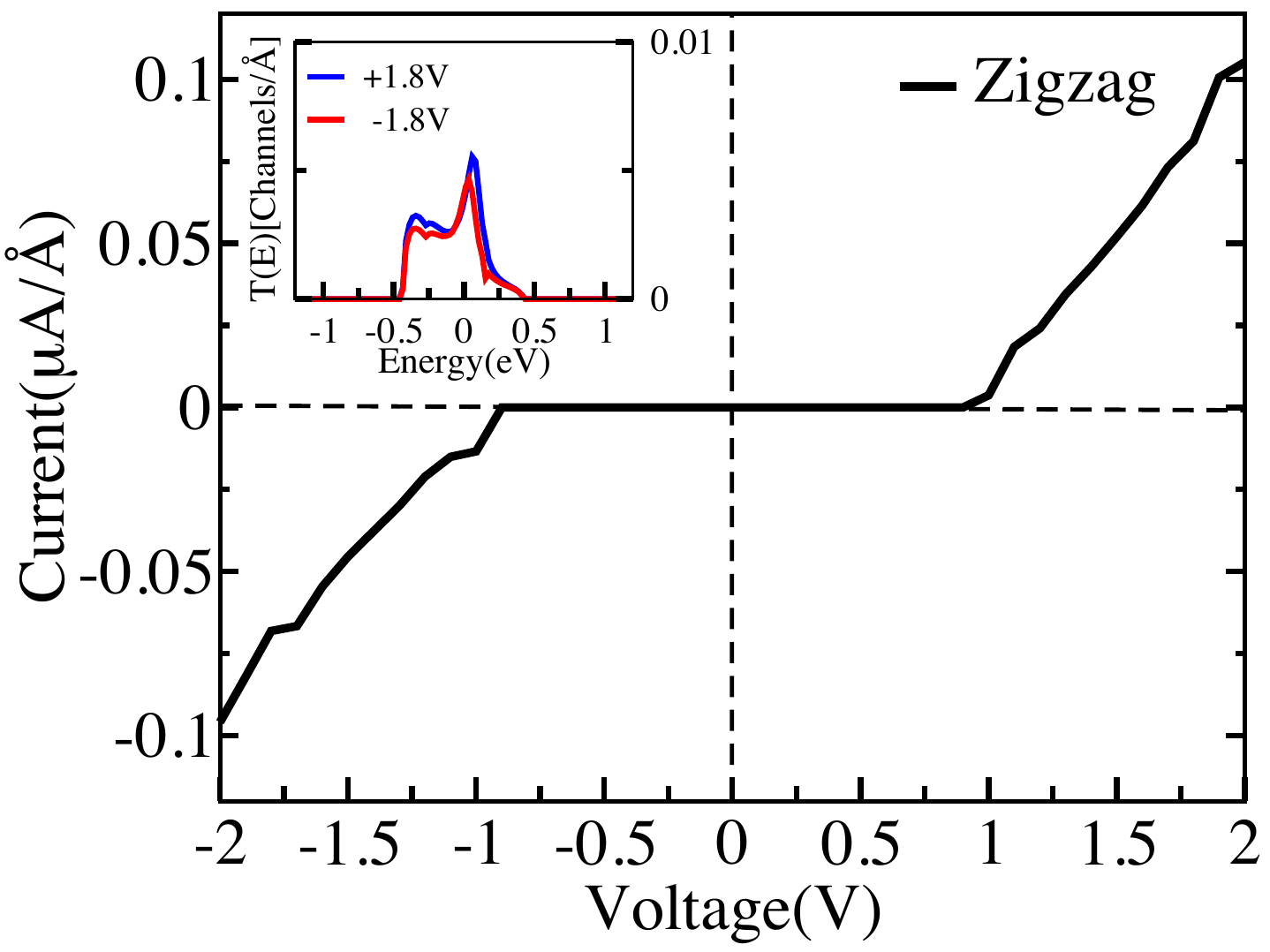}}\
\subfloat[]{\includegraphics[width = 5cm,angle=270]{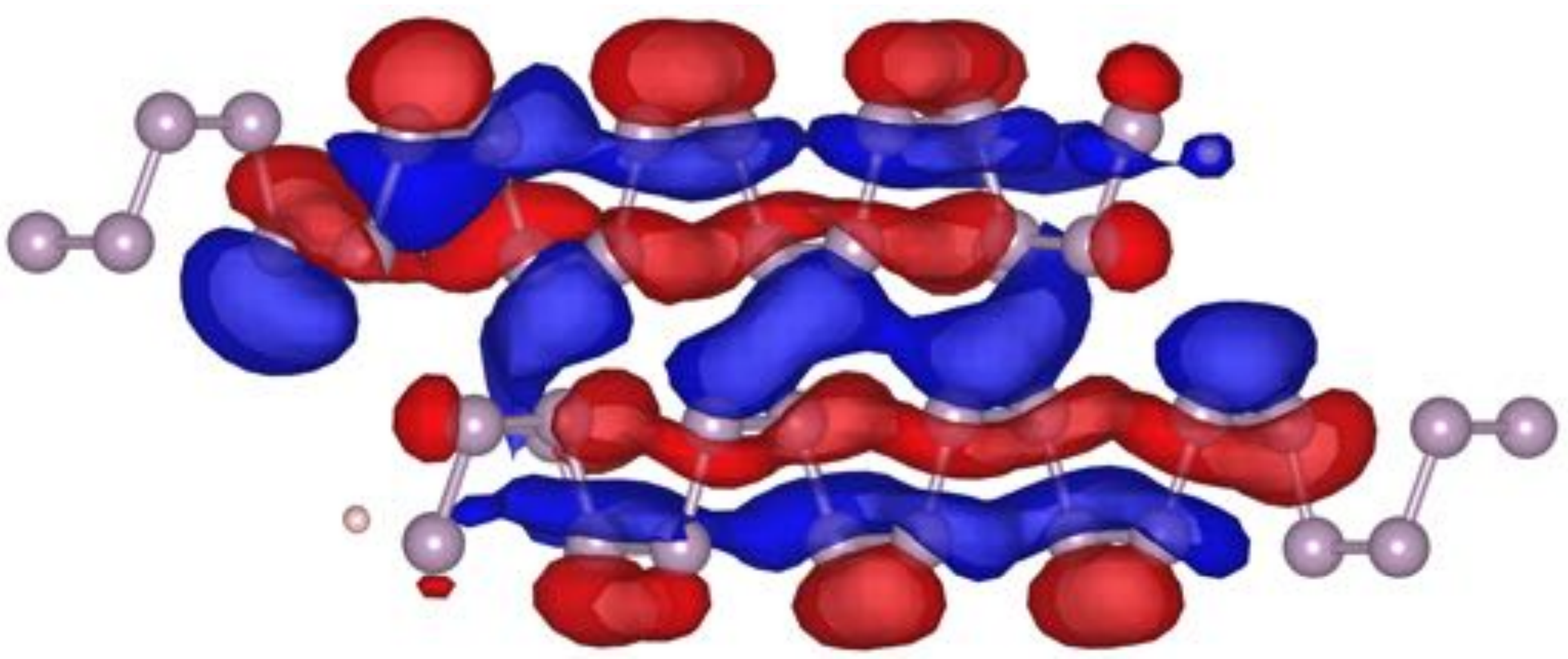}}
\subfloat[]{\includegraphics[width = 5cm,angle=270]{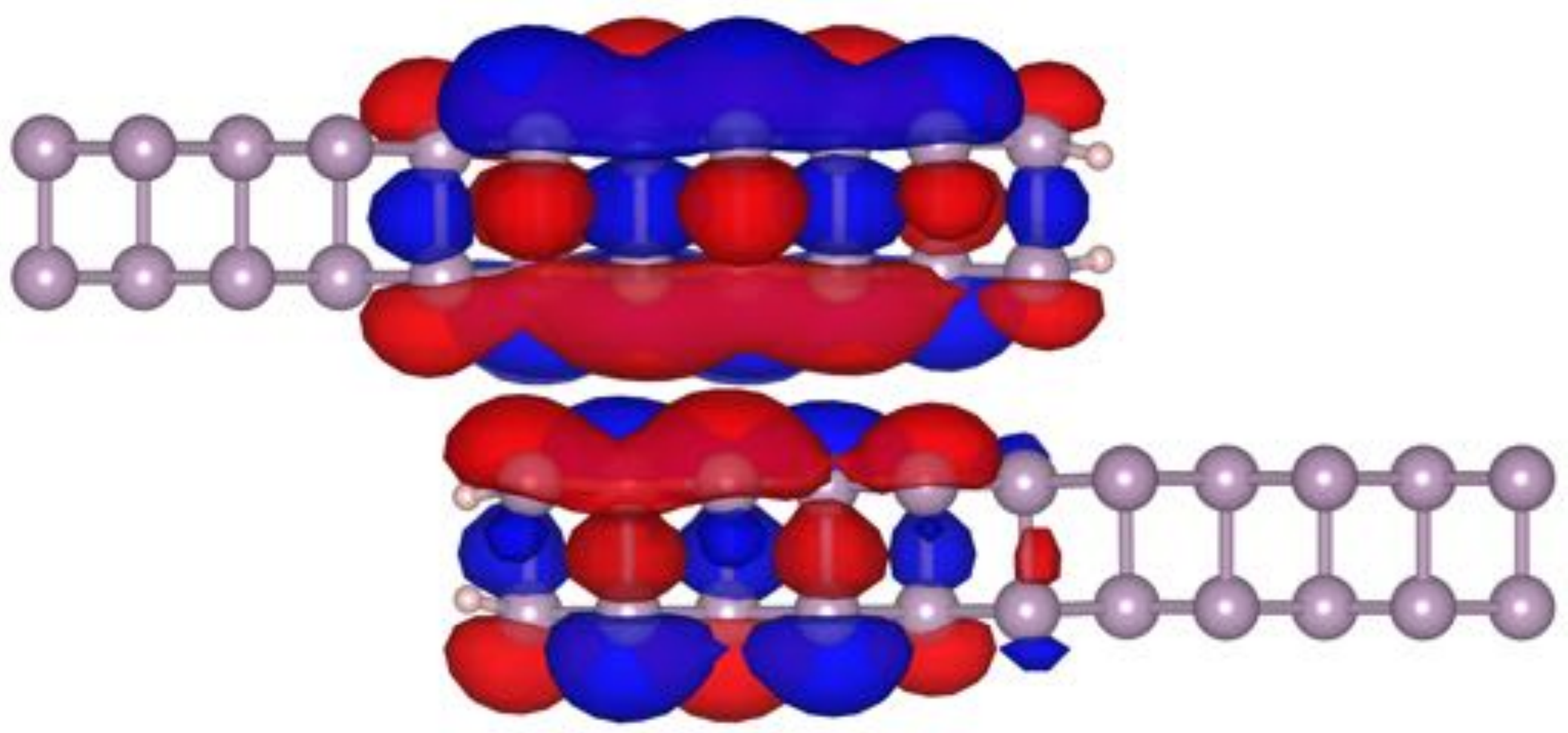}}
\caption{Current-Voltage characteristics for symmetric nanojucntion (a) Armchair (b) Zigzag direction (Insets show the transmission function at ±1.8V bias in respective device) and eigenchannels for rotated bilayer nanojunctions at +1.8V at 0 eV (c) Armchair direction (d) Zigzag direction}
\end{figure}

	To understand the interaction between the layers in bilayer nanojunction region and its influence upon the current-voltage (I-V) characteristics, we increased the bias in periodic order of 0.1 V/step and calculated the current using the Landauer formula. Initially the current remains nearly zero but when it overcomes the energy gap of the electrodes it gives the symmetric IV characteristics for both the positive and negative biases in Fig. 3(a). This symmetric IV comes due to the peak in the transmission in the bias window at increased bias, see inset of Fig. 3(a). We see fano resonance behavior in the transmission at the 0 eV level at 1.8 V bias. We also tried to decrease the junction area, this eliminates the fano resonance behavior while the overall current behavior does not change.

	We now consider the case on the zigzag direction which has lower transmission than that of armchair direction at zero bias in Figure 4(b). This clearly demonstrates the anisotropic behavior of BP in such a device. Monolayer BP is expected to have 100 times lower conductance than armchair direction\cite{kou2014phosphorene} and it reduces further when it comes to bilayer form. Current in zigzag direction is also symmetric with respect to applied voltage, as it can be seen in Figure 3(b).

	The transmission function at ±1.8 V can be explored in inset of the fig. 3(b). Overall transmission in voltage window range has much less magnitude than that of armchair junction. The transmission peak yields the transmission coefficient of 0.1 channels/\AA{} in the armchair direction whereas in the zigzag direction it comes out to be only 0.005 Channels/\AA{}, and lower transmission value implies lower current. The anisotropy factor (I$_arm$/I$_zig$) in current in these two nanojunctions, which are in mutually perpendicular direction is of around 40 at 2 eV bias. This anisotropy is lower than the reported value for the monolayer form of BP. Further, we do not see current rectification in these two devices and it is quite expected due to the symmetric structure of the device. The eigenchannels for the two devices at 1.8 V are рlotted at the center of voltage window. Fig. 3(c), presents the eigenchannel for armchair direction, which argues that there are conductive states spread all over the junction area and also in between the layers. This implies better interlayer coupling and higher interlayer tunneling probability. There are no significant edge state contributions in the eigenchannel of the armchair device. On the other hand good coupling at 1.8 V bias implies, that conduction band of the left electrode couples to the valence band of the right one, but also an induced state is created within the bilayer area, which is responsible for the Fano resonance in transmission. This induced state naturally disappears when contact area vanishes. 
 
\begin{figure}
\centering
\subfloat[]{\includegraphics[height=2.6cm,width = 8.4cm]{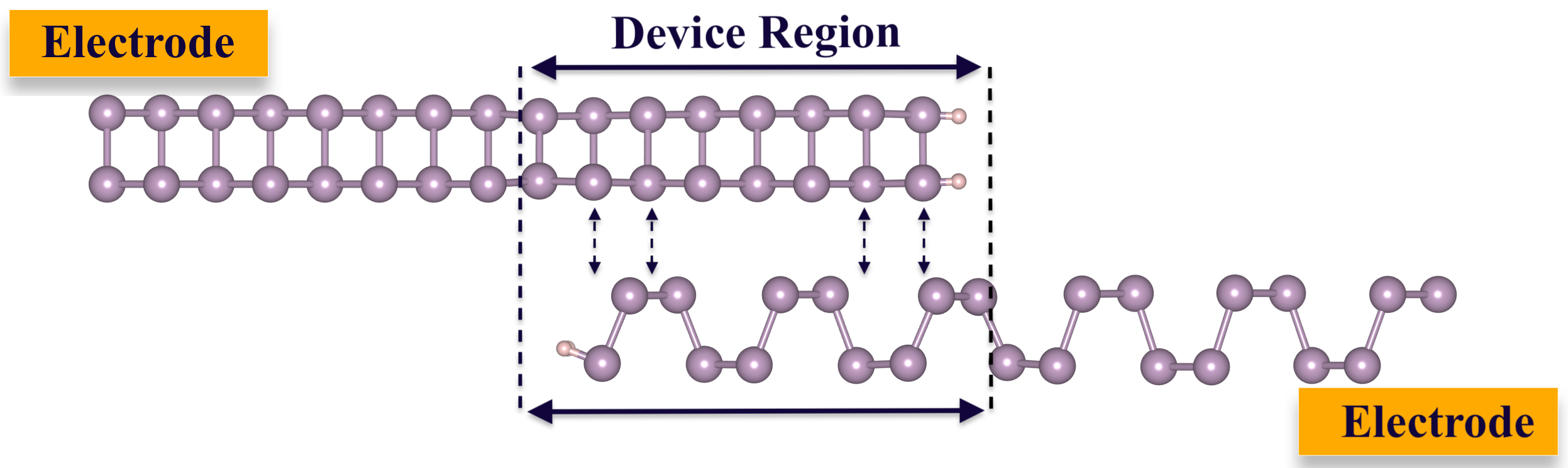}}\
\centering
\subfloat[]{\includegraphics[width = 5cm]{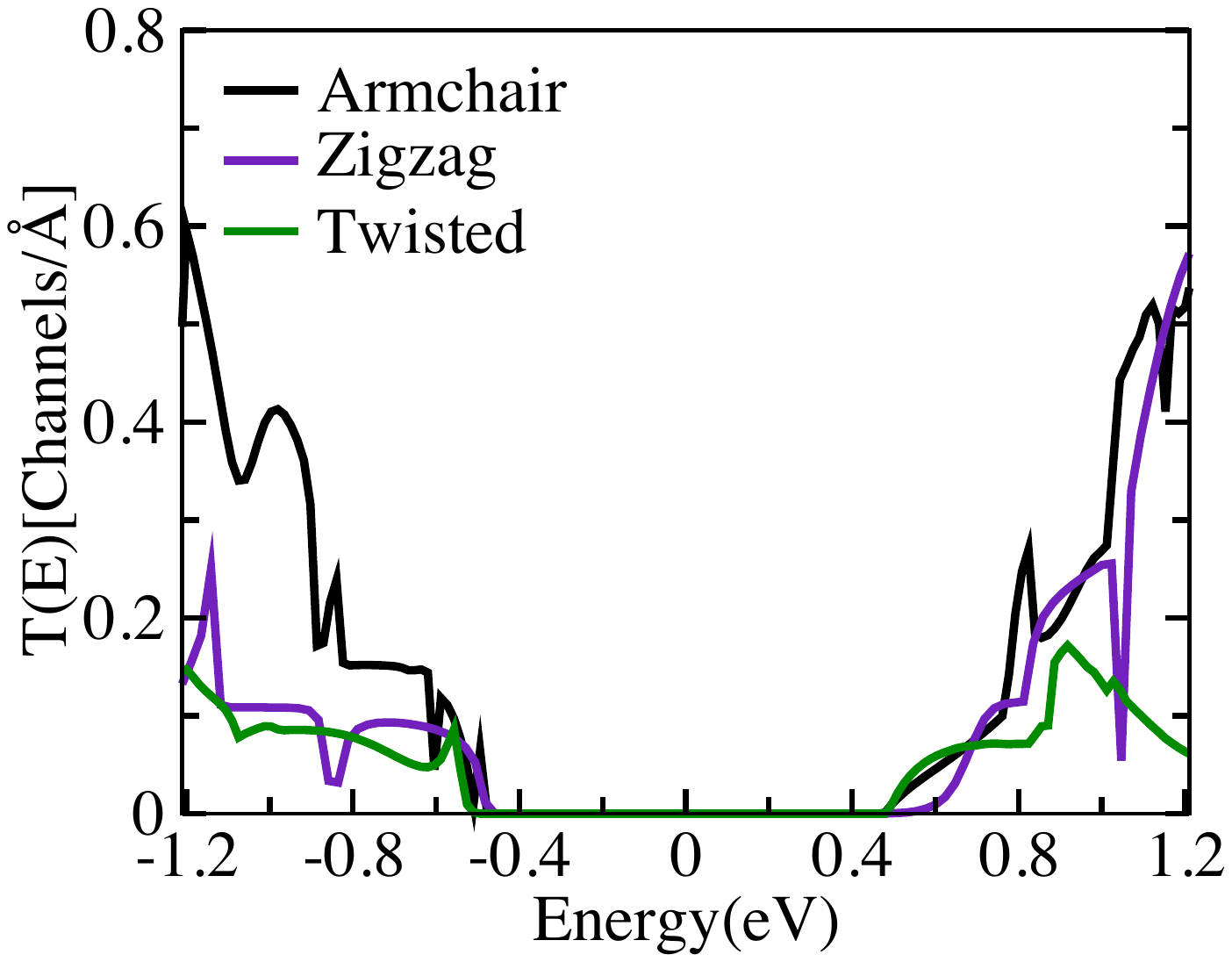}}
\subfloat[]{\includegraphics[width = 5cm]{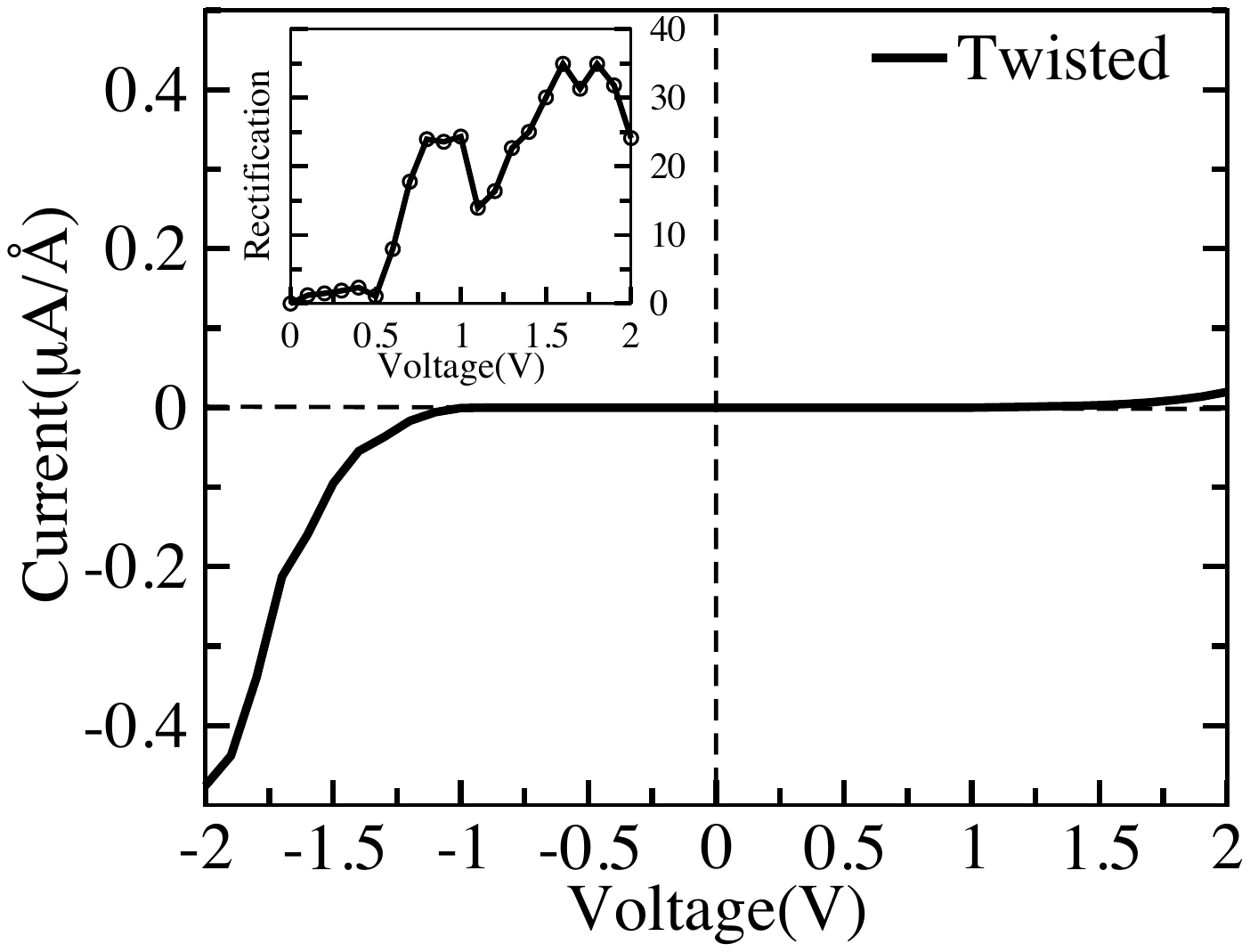}}
\caption{(a) Schematic picture for twisted bilayer nanojunction ( device reagion has bilayer of armchair and zigzag direction) (b) Zero bias transmission for the nanojunctions (c) Current-Voltage characteristics for twisted nanojunction device (Inset shows the rectification ratio of the respective device)}
\end{figure}

	For the junction in in the zigzag direction, 1.8V bias eigenchannel is localized all over the junction but it gets more localized at the lower layer, where it does not connect to the right electrode (last 4 rows of atoms at Fig. 3(d). Also there are no states connecting the layers, so it reduces the probability of electron tunneling from one layer to the other. This explains the lower current in the zigzag direction. In fact, the eigenchannel at Fig. 3(d) looks qualitatively same as eigenchannel for the Fano state looks in armchair direction: we see an incoming state at the upper BP sheet, and the induced state below. However, because the conducting state that couples the layers is not present on this junction, the state generate not a Fano, but Wigner-shaped tunneling resonance, which is clearly seen in the inset of the Fig. 3(b) at 0 eV.

	There are possibilities to tune the anisotropy factor (I$_arm$/I$_zig$) by increasing the junction, which will imply a bigger quantum well, as explained by Zhang et. al \citep{zhang2018quantum}. The levels in the well can tune the resonate tunneling of the electrons in-between the layers which can further tune the anisotropy factor of the system. 

	To further take the full advantage of anisotropic behavior of BP, we introduce a twisted nanojunction in Figure 4(a), where the band structure and DOS of the twisted layer have already been explained above. Here, we again have the ML-BL-ML architecture but one of the monolayers is in armchair direction and the other has zigzag direction. The central region has a structure of a vertically stacked 90\textdegree twisted bilayer, which would be isotropic, unless now we have twisted nanojunction setup with semi-infinite electrodes in either directions. The upper layer has zigzag transport direction and lower layer has armchair transport direction, and both layers have one edge terminated with hydrogen. In our case of study we took junction length of 12\AA{}. The zero bias transmission in Fig. 4(b), shows that the overall transmission is lower than that of the armchair and zigzag directions. This is due to the twisted layer, which has less interaction between the BP layers than that of the AA-stacked bilayer. The transmission gap remains nearly the same as for zigzag and armchair nanojunction, which, in the case of the nanojunction device, is simply the gap of the monolayer BP electrodes. 
 
	Asymmetry is expected to bring in some rectification behavior; to check this, we increased the bias symmetrically over the junction with the step of 0.1V. We contemplate a behavior in the I-V characteristics in Fig 4(c), where it shows current rectifying property. When the electrons are injected from armchair to zigzag direction in negative bias regime, the current is higher, whereas when it is injected from zigzag to armchair direction, the current is lower. Inset of Fig. 4(c), shows rectification ratio (I$_-$/I$_+$) with maximum of 35 achieved at 1.8 V bias. 

	To understand the rectification behavior of the nanojunction, we plot the transmission function at ±1.8 V bias in Fig. 5(a), demonstrating high transmission probability for negative bias (-1.8 V) and much less transmission for positive bias (+1.8 V) in the bias window. In the negative bias, we notice the transmission at the 0 eV but also higher peak comes at the 0.35 eV. Further we plot the eigenchannels for both the biases as ±1.8 V for 0 eV energy in Fig. 5(b-c). It is quite obvious from the eigenchannels that for the negative bias there is strong tunneling probability from armchair to zigzag layer as the channel is spread all over the device region which implies the higher conductance. In the positive bias (+1.8V), the channel is more localized towards the left side of the nanojunction and remains only in the zigzag layer, which yields lower tunneling from zigzag to armchair layer and provides the lower conductance. We have already discussed in fig. 4(c) that in the negative bias, higher transmission peak comes at +0.35 eV, henceforth we also plotted the eigenchannel at this energy in Figure 5(d). This presents a strong tunneling channel, delocalized all over the nanojunction area and responsible for much higher current in the negative bias regime. 

	Role of the edge states can be estimated from results where the length (and the area) of the BP sheets overlap was decreased, yield lower rectification ratio, see Supplementary information Fig. S2. In this case the edge states take significant part in conductance mechanism, while rectification is mainly due to the interlayer tunneling. Increase of the junction length means bigger quantum well and bigger length for interlayer tunneling which implies the higher rectification up to certain limiting length. The length dependent rectifying property can be the case of further study. Also the termination of edge can play an important role in defining the rectifying property. Another limitation of our work is, that we use BP as a lead to make simulations reasonably less expensive. In experiment one would inject electrones from metallic electrodes, and also control the electrodes' potentials through metal. We model such situation via directly controlling the electrodes, the limitation is that we miss metal induced states, which might slightly affect the rectification ratio. This also suggests yet another method of controlling rectification by placing the junction asymmetrically with respect to the electrodes, which we do not explore. 
	
	It is very interesting to see the rectifying behavior in a device where no electron transfer between the layers and no band bending exist. It means there is no standard p-n junction type of characteristics preset in the beginning. Comparing the I-V characteristics of rotated nanojunction with the armchair and zigzag nanojunction, it is evident, that for the positive bias, when the electron flow is from zigzag to armchair direction the limiting current regime is zigzag directional current. When current flow is from armchair to zigzag direction it follows the armchair nanojunction for limiting current.

	Twisted bilayer has been reported for gate switchable transport anisotropy with applied gate voltage. We also report the gate tunable electronic structure of the twisted bilayer in Fig. S4, where the out of plane gate voltage can tune the isotropic behavior of band twisted bilayer in periodic form. Henceforth, we suggest that in the real time device transport behavior can be controlled by the applied gate voltage, which further opens the wide range of possibility to tune the current and even yield much higher rectification which makes the setup desirable in nano-electronic applications. 

\begin{figure}[h]
\centering
\subfloat[]{\includegraphics[width = 5cm]{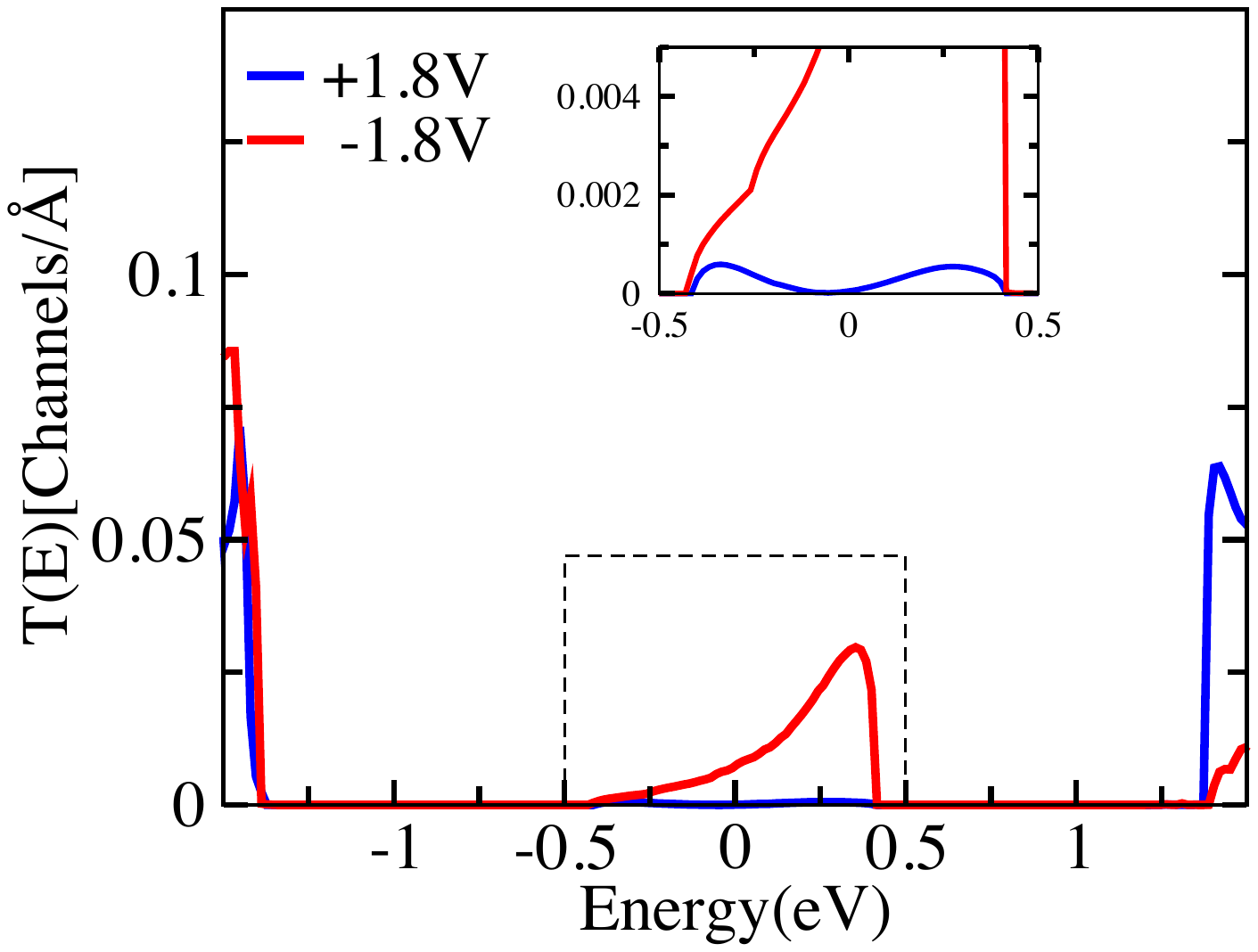}}\
\subfloat[]{\includegraphics[width = 4.5cm, angle=270]{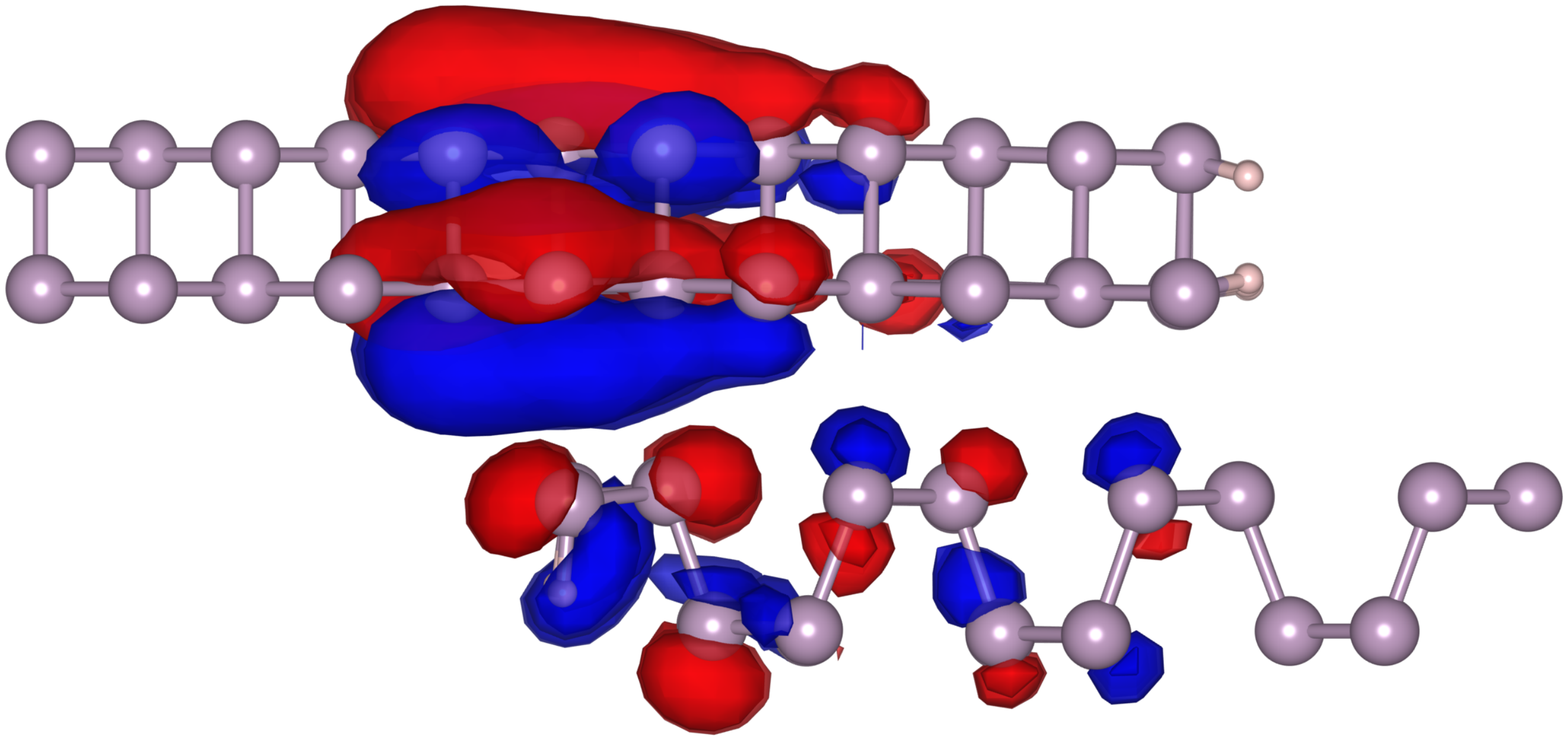}}
\subfloat[]{\includegraphics[width = 4.5cm, angle=270]{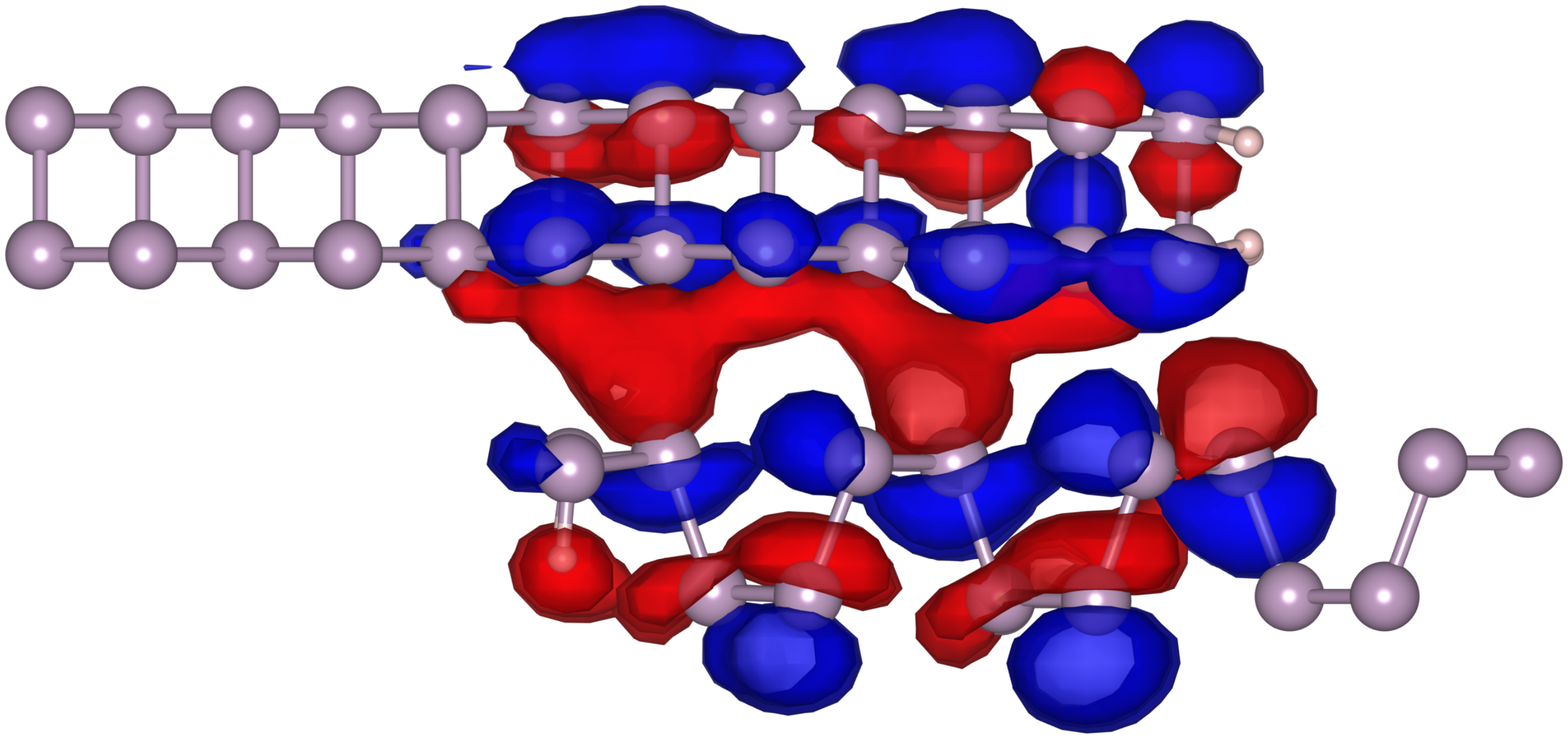}}
\subfloat[]{\includegraphics[width = 4.2cm, angle=270]{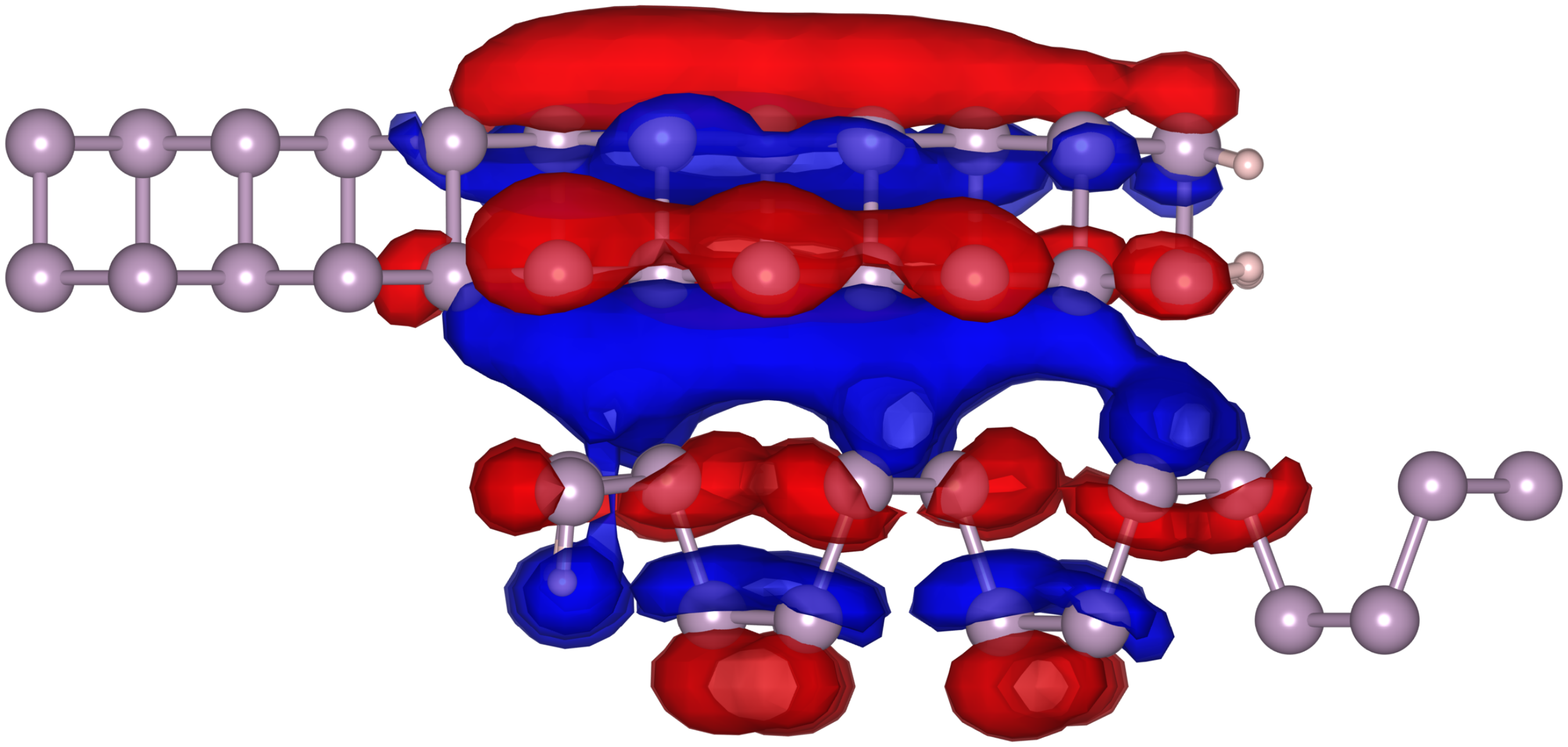}}
\caption{ (a) Transmission at ±1.8 V bias for twisted bilayer nanojunction (inset shows the zoomed picture of transmission around 0 eV) and eigenchannels for rotated bilayer nanojunction at 0 eV (a) at positive bias of 1.8 V (b) negative bias of 1.8V and (c) eigenchannel at 0.35 eV at -1.8 V bias}
\end{figure}

	In conclusion by using DFT along with NEGF calculations, we report a unique nanoscale rectifying device which consist of two phosphorene sheets in the parallel contact with 90\textdegree twisted angle without anything additionally bridging the electrodes. This rectifier has advantages over the other 2D heterostructure-based p-n junctions where one has to care about the band alignment and contact region. We also studied two symmetric nanojunctions oriented along armchair and zigzag directions with different resistivities due to the anisotropy of the BP. We investigate the rectifying property by using eignechannels and their localization across the junction. No current rectification is observed in symmetric junctions, yet resistivities are different along different directions. In asymmetric 90\textdegree twisted junction electron flow along "easy" direction is spread over the junction area, electrons tunnel between overlapping layers, but also induced state is seen as Fano resonance. Along the opposite direction electrons are mostly reflected from the edge of the sheet, tunneling of electron from one layer to the other is inefficient and induced states start to play frontmost role in transport. Directional anisotropy is the only the reason behind this rectifying behavior of the device. Many tuning mechanisms can be aforeseen for BP device, like perpendicular electric field (through gate voltage), strain, asymmetric design and variable contact area. Our work further opens the possibilities for implementing field effect transistor or rectifier. This is to our knowledge the first time when anisotropy in the system has been proposed to produce the rectifying behavior in nanojunction device. 

\subsection{Method}

First principles density functional theory (DFT)\cite{hohenberg1964inhomogeneous, kohn1965self} calculations were performed using SIESTA\cite{soler2002siesta, artacho2008siesta} program within generalized gradient approximation with Perdew, Bruke and Ernazerholf (PBE) \cite{perdew1996generalized} functional for geometrical optimization through total energy calculation. Norm-conserved Troullier-Martins pseudopotentials were used to describe the interaction between core and valence electrons \cite{troullier1991efficient}. The mesh cut off was 200 eV and Brillouin zone integration for the supercell was sampled by 2x2x1 k-points, and for unit cell it was 20x30x1 within Monkhorst-pack scheme with double-polarized basis set \cite{monkhorst1976special}. To simulate the unit and super cell of bilayers periodic boundary conditions were used with 30\AA{} vacuum space to minimize the interaction between the layers. The systems were fully relaxed to obtain the ground state structure with residual forces on the atoms less than 0.01 eV/atom. The quantum transport properties have been studied combining non-equilibrium Green’s function (NEGF) and DFT in TranSIESTA \cite{brandbyge2002density} module using Gamma k-points grid which yields reasonably converged results in comparison 1x4 k-point grid. The transmission spectrum, which defines the probability for electrons to be transferred from left to right electrode with the specific energy E, is calculated from the equation \cite{datta1997electronic},

\begin{equation}
 T(E,V) = tr[\Gamma_R(E,V)G_C(E,V)\Gamma_L(E,V)G_Cˆ{a}(E,V)]
\end{equation}

Where $G_C$ is the Green’s function of the central region and is the coupling matrix of electrodes in either sides. The integration of this transmission function gives the electric current, 
\begin{equation}
I(V) = \int_{\mu_L}^{\mu_R} T(E,V) \Big\{f(E-{\mu_L})-f(E-{\mu_R})\Big\}dE
\end{equation}
Where $\mu_L$ = -V/2 ($\mu_R$=V/2) is the chemical potential of the left and right electrode.

\begin{acknowledgement}
The authors acknowledge computational resources provided through Swedish National Infrastructure for Computing (SNIC2017-11-28, SNIC2017-5-8). VS acknowledges funding from the European Erasmus fellowship program. AG and RA acknowledge support from the Swedish Research Council.
\end{acknowledgement}

\begin{suppinfo}

This will usually read something like: ``Experimental procedures and
characterization data for all new compounds. The class will
automatically add a sentence pointing to the information on-line:

\end{suppinfo}



\bibliography{achemso-demo}

\end{document}